\documentclass[fleqn,usenatbib]{mnras}

\usepackage{newtxtext,newtxmath}

\usepackage[T1]{fontenc}

\DeclareRobustCommand{\VAN}[3]{#2}
\let\VANthebibliography\thebibliography
\def\thebibliography{\DeclareRobustCommand{\VAN}[3]{##3}\VANthebibliography}

\usepackage{graphicx}	
\usepackage{amsmath}	
 \usepackage{xcolor}

\def\civ{\ion{C}{IV}}
\def\caii{\ion{Ca}{II}}
\def\heii{\ion{He}{II}}
\def\hii{\ion{H}{II}}
\def\nv{\ion{N}{V}}
\def\nai{\ion{Na}{I}}
\def\nevii{\ion{Ne}{VII}}
\def\neviii{\ion{Ne}{VIII}}
\def\oii{\ion{O}{II}}
\def\oiii{\ion{O}{III}}
\def\ov{\ion{O}{V}}
\def\ovi{\ion{O}{VI}}
\def\ovvi{\ion{O}{V-VI}}

\def\Gaia{{\it Gaia}}

\newcommand{\Teff}{\mbox{$T_\mathrm{eff}$}}

\title[\ion{O}{VI} Nuclei of PNe]{Spectroscopic survey of faint planetary-nebula nuclei. I\null. Six new ``\ion{O}{VI}'' central stars\thanks{Dedicated to the memory of James B. Kaler (1938 December 29--2022 November 26): kind friend and colleague, pioneer in the modern study of planetary nebulae and their central stars.}}

\author[H. E. Bond et al.]{
Howard E. Bond,$^{1,2}$\thanks{E-mail: heb11@psu.edu}
Klaus Werner,$^{3}$
George H. Jacoby,$^{4}$ 
and Gregory R. Zeimann$^{5}$
\\
$^{1}$Department of Astronomy \& Astrophysics, Pennsylvania State University, University Park, PA 16802, USA\\
$^{2}$Space Telescope Science Institute, 3700 San Martin Dr., Baltimore, MD 21218, USA\\
$^{3}$Institut f\"ur Astronomie und Astrophysik, Kepler Center for Astro and Particle Physics, Eberhard Karls Universit\"at, Sand 1, D-72076 T\"ubingen, Germany \\
$^{4}$NSF's NOIRLab, 950 N. Cherry Ave., Tucson, AZ 85719, USA \\
$^{5}$Hobby-Eberly Telescope, University of Texas, Austin, Austin, TX 78712, USA
}

\date{Accepted XXX. Received YYY; in original form ZZZ}

\pubyear{2023}

\begin{document}
\label{firstpage}
\pagerange{\pageref{firstpage}--\pageref{lastpage}}
\maketitle

\begin{abstract}
We report initial results from an ongoing spectroscopic survey of central stars of faint planetary nebulae (PNe), obtained with the Low-Resolution Spectrograph on the Hobby-Eberly Telescope. The six PN nuclei (PNNi) discussed here all have strong emission at the \ovi\ 3811--3834~\AA\ doublet, indicative of very high temperatures. Five of them---the nuclei of Ou~2, Kn~61, Kn~15, Abell~72, and Kn~130---belong to the hydrogen-deficient PG\,1159 class, showing a strong absorption feature of \heii\ and \civ\ at 4650--4690~\AA. Based on exploratory comparisons with synthetic model-atmosphere spectra, and the presence of \neviii\ emission lines, we estimate them to have effective temperatures of order 170,000~K. The central star of Kn~15 has a Wolf-Rayet-like spectrum, with strong and broad emission lines of \heii, \civ, \nv, and \ion{O}{V-VI}. We classify it [WO2], but we note that the \nv\ 4604--4620~\AA\ emission doublet is extremely strong, indicating a relatively high nitrogen abundance. Several of the emission lines in Kn~15 vary in equivalent width by factors as large as 1.5 among our four observations from 2019 to 2022, implying significant variations in the stellar mass-loss rate. We encourage spectroscopic monitoring. Follow-up high-time-resolution photometry of these stars would be of interest, given the large fraction of pulsating variables seen among PG\,1159 and [WO] PNNi.  
\end{abstract}

\begin{keywords}
stars: emission-line, Be -- stars: winds, outflows --  stars: Wolf-Rayet --  planetary nebulae: general -- white dwarfs
\end{keywords}



\section{Introduction: Nuclei of Planetary Nebulae}

Planetary nebulae (PNe) mark the transition of an asymptotic-giant-branch (AGB) star across the HR diagram to the
onset of the white-dwarf (WD) cooling sequence. At the end of the AGB stage, the
outer layers of the star (initial masses in the range $\sim$0.8--$8\,M_\odot$)
are ejected, and the remnant core rapidly evolves to higher temperature. When the star
reaches an effective temperature of $\sim$30,000~K, its ultraviolet (UV) radiation ionizes the surrounding ejecta,
producing the PN---which lasts a few times $10^4$~yr before 
dissipating. For a recent review of PNe, see \citet{Kwitter2022}.
The central stars of PNe---planetary-nebula nuclei (PNNi)---show a remarkable
range of spectroscopic and photometric phenomena. These include the effects of binary companions, unusual chemical compositions, pulsations, and stellar winds.

In recent years, deep-sky surveys have revealed numerous new low-surface-brightness PNe.
An example is the visual search of Digitized Sky Survey (DSS) 
images by the ``Deep-Sky Hunters'' (DSH) collaboration. The DSH are a group of mostly amateur astronomers who initially were searching for previously unknown open star clusters, but as a byproduct discovered about two dozen PN candidates \citep{Kronberger2006}. Within a few years, the DSH team had found about 100 new candidate PNe; see \citet{Jacoby2010}, where the search techniques are described. Six years later, the number of PN candidates from the DSH survey had grown to more than 250 \citep{Kronberger2016}. Meantime, a large group of amateurs, located primarily in France, has discovered more than 200 new PNe, plus hundreds of further candidates \citep{LeDu2022}, using a variety of techniques including long-exposure CCD and CMOS imaging with small telescopes. A website\footnote{\url{ https://planetarynebulae.net/EN/index.php}} maintained by P.~Le~D\^u and T.~Petit gives details about these amateur discoveries. Extensive information about individual PNe, including finding charts and multi-wavelength images, is contained in the Hong-Kong/AAO/Strasbourg/H$\alpha$ Planetary 
Nebulae (HASH) database\footnote{\url{http://hashpn.space/}} \citep{Parker2016}. 

Many of the newly discovered PNe have faint, and often blue, stars near their centres, but relatively little is known about these candidate PNNi. 
In response to these discoveries, in 2019 H.E.B. began a spectroscopic survey of these potentially new central stars. 
The survey is modeled on the early work of \citet{Napiwotzki1995}, who obtained and classified spectra of $\sim$3~dozen faint PNNi. In a subsequent followup, about 200 spectral types for PNNi were assembled from the literature by \citet{Napiwotzki1999}. More recently, \citet{Weidmann2020} have assembled a catalogue of spectral types for nearly 600 PNNi from the literature.

The aim of the present survey is to obtain spectra of modest resolution and signal-to-noise ratio (SNR), for the purposes of spectral classification, and to identify objects worthy of followup at higher resolution for more detailed analysis. In this first paper from the survey, we present results for a half-dozen especially interesting central stars. These objects are members of the rare class of extremely hot central stars with strong emission lines of \ovi, described in detail below. In subsequent papers, we will present spectral types for several dozen additional objects.

\section{Observations}

Spectroscopic observations for the survey are obtained with the second-generation Low-Resolution Spectrograph (LRS2) of the 10-m Hobby-Eberly Telescope (HET; \citealt{Ramsey1998,Hill2021}), located at McDonald Observatory in west Texas, USA\null. The LRS2 instrument is described in detail by \citet{Chonis2014, Chonis2016}, but we give an overview here. 

LRS2 provides integral-field-unit (IFU) spectroscopy with 280 $0\farcs6$-diameter lenslets that cover a $12''\times6''$ field of view (FOV) on the sky. LRS2 is composed of two arms: blue (LRS2-B) and red (LRS2-R). All of our observations are made with the target placed in the LRS2-B FOV\null. The LRS2-B arm employs a dichroic beamsplitter to send light simultaneously into two spectrograph units: the ``UV'' channel (covering 3640--4645~\AA\ at resolving power 1910), and the ``Orange'' channel (covering 4635--6950~\AA\ at resolving power 1140).


HET observations are carried out in a queue-scheduling mode \citep[see][]{Shetrone2007PASP}. { We submitted a lengthy list of candidate targets, selected primarily from the sources described in \S1. Our PNNi survey belongs to the lowest priority level (P4) in the queue, meaning that data are obtained when schedule gaps and/or marginal weather or seeing conditions preclude higher-priority observations. Thus the observed targets are effectively chosen at random from the input list.} We generally obtain several observations of each target, allowing the data for each object to be combined, increasing the SNR of the final spectrum. Moreover, multiple observations allow us to search for spectral changes, such as those due to stellar-wind variations, binary companions, or spectacular mass-loss events such as those seen in the central star of Lo~4 \citep{Bond2014}.

\section{Data Reduction}

Raw LRS2 data are processed with \texttt{Panacea}\footnote{\url{https://github.com/grzeimann/Panacea}}, an automated reduction pipeline written by G.R.Z. (Zeimann et al., in preparation).  Initial processing includes bias and flat-field correction and wavelength calibration. \texttt{Panacea} uses a spatial and spectral point-spread-function filter to identify the highest-SNR object separately in each channel (UV and Orange). In this paper we discuss the spectra of six central stars, described in detail below. For four of the six targets, the brightest object in the field was our desired central star; however, for the objects Kn~12 and Kn~15 there was a neighboring and cooler companion that the automatic pipeline selected instead in the Orange channel.  We discuss those cases in the next paragraph. For the other objects, after the target is identified, sky fibers are selected to be $>$$3''$ from the source, and a median sky spectrum is constructed.  This generally includes emission from the surrounding PN, frequently resulting in an over- or under-subtraction in the stellar spectrum at the wavelengths of nebular lines, such as [\oii] 3727~\AA, [\oiii] 4959--5007~\AA, and the Balmer series. Thus the LRS2-B stellar spectra often provide little useful information at, for example,  H$\alpha$ and H$\beta$. After sky subtraction, the stellar spectrum is extracted by fitting a 2D Gaussian to the highest-SNR image, collapsed in the dispersion direction. Flux calibration is based on default response curves generated from standard stars observed over many nights, measures of the mirror illumination, and the exposure throughput from guider images.  



\begin{figure}
\centering
\includegraphics[width=0.47\textwidth]{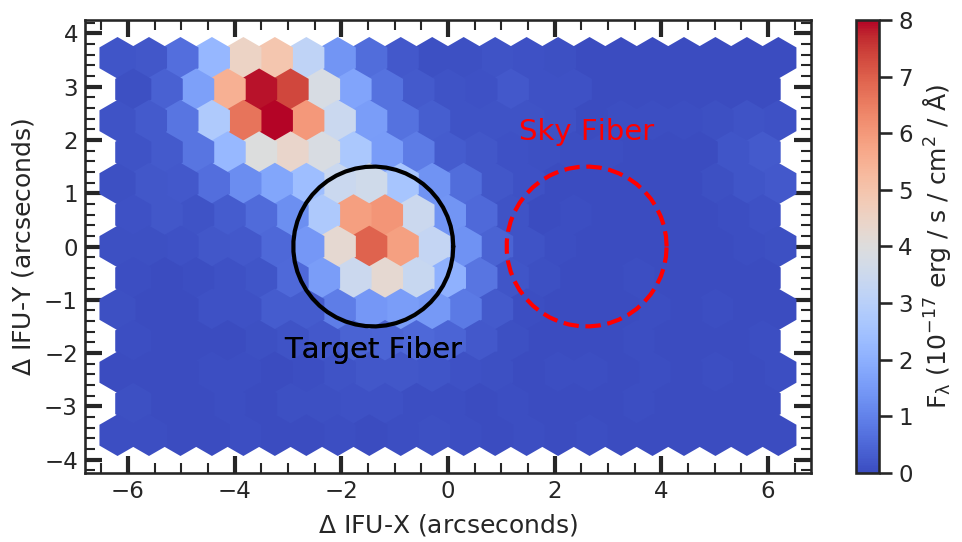}
\caption{Artificial ``image'' of the Kn~15 field in the LRS2-B IFU plane for our observation on 2020-11-10. { North is to the right, east toward the top.} The spectra from each fiber are collapsed along the dispersion direction, and we show the signal from the stellar \civ\ emission doublet in the range $5800 \pm 10$~\AA, colour-coded according to the mean flux density in units of $10^{-17}\rm\, erg\, s^{-1}\, cm^{-2}\,$~\AA$^{-1}$.  The  extraction aperture for Kn~15 with radius $1\farcs5$ is marked with a black circle { labelled ``Target Fiber,'' } and the fixed ``sky'' aperture with a red circle, located 4$\arcsec$ away (the furthest distance we could go and still stay in the field of view for all observations). A bright ($G=16.2$) field star lies to the top left of the Kn~15 central star, at a separation of $2\farcs96$ { and a position angle relative to north of $123^\circ$}.
\label{fig:kn15ifu} }
\end{figure}

As noted, Kn~12 and Kn~15 posed special problems because of nearby field stars falling within the LRS2-B FOV\null. After the initial reduction, we used
\texttt{LRS2Multi}\footnote{\url{https://github.com/grzeimann/LRS2Multi}}, a \texttt{python} interface, to perform advanced reduction steps and calibrations for \texttt{Panacea} products.  We manually identified the target PNN in each exposure, and defined extraction apertures with radii of $2\farcs5$ and $1\farcs5$ for Kn~12 and Kn~15, respectively.  Based on the configuration of the neighboring stars, we used fibers beyond $3\farcs5$ for Kn~12 to build our sky model for each exposure.  In the case of Kn~15, there is a bright star lying only $2\farcs96$ from the PNN\null. Here we defined a fixed sky aperture  $4\farcs0$ away, with the same $1\farcs5$ extraction radius as for the target, as shown in Figure~\ref{fig:kn15ifu}.  This was done in order to subtract the nebular emission consistently from all of the target spectra.  For the final spectra of all of the targets, we masked the strongest PN emission lines, as well as the night-sky emission lines at 5577~\AA\ and 6300~\AA\null. We used a $\pm$6~\AA\ window and 6~\AA\ Gaussian kernel to interpolate over the masked regions.

\section{``\ovi'' Planetary Nuclei}

The majority of PNNi have hydrogen-rich atmospheres, but about one-third are hydrogen-deficient (see, for example, the \citealt{Weidmann2020} catalogue). Among hydrogen-poor central stars, a small number exhibit strong emission lines of the \ovi\ 3811--3834~\AA\ doublet, thus placing them among the hottest known stars. They were assigned to an ``\ovi'' spectral type in early PNN classification schemes \citep[e.g.,][]{Smith1969, Aller1975, Heap1975, Heap1982, Kaler1984}. \citet{Kaler1991} recognized two subtypes of \ovi\ central stars: those in which the \ovi\ doublet is well resolved, and those in which the \ovi\ emission comes from a stellar wind having such a high velocity that the doublet is blended into a single strong, broad feature at $\sim$3820~\AA.  

Hydrogen-deficient PNNi are generally considered to belong to an evolutionary sequence of evolved post-AGB stars that have experienced a late or very late thermal pulse. The re-ignition of helium shell burning returns the star to an earlier stage of low temperature and large radius---a ``born-again'' red giant---and in the process the surface hydrogen is burned. The H-deficient star then retraces its previous post-AGB evolution, proceeding first to higher temperatures, and eventually beginning its descent of the WD cooling sequence. As the star evolves, an initially dense stellar wind gradually diminishes and fades away. See \citet{Mendez1986}, \citet{Werner2003}, \citet{WernerHerwig2006}, \citet{Lobling2019}, \citet{Weidmann2020}, and references therein, for discussions of such evolutionary scenarios. 

Post-thermal-pulse central stars with dense, hydrogen-deficient stellar winds have spectra that resemble those of massive carbon-type Wolf-Rayet (WR) stars. In modern spectral-classification terminology, such PNNi are assigned to spectral type [WC], the brackets distinguishing them from massive WR stars. These stars can have spectral types as late as [WC11] or [WC12] \citep[e.g.,][]{Margon2020}. Their subsequent evolution takes them through earlier [WC] types, to extremely hot [WO] stars (an alternative terminology for \ovi\ nuclei), and thence to the ``PG~1159'' hot WDs.  See, for example, Figure~1 in \citet{WernerHerwig2006} and Figure~2 in \citet{Weidmann2020} for schematic illustrations of this ``classical'' evolutionary sequence of spectral types.
The PG~1159 WDs are, in turn, the likely progenitors of the cooling sequence of hydrogen-deficient DO, DB, and DQ WDs \citep[e.g.,][and references therein]{Bedard2022}.



{

The spectra of PG~1159 WDs are characterized by a distinctive blend of \heii\ and \civ\ features in an absorption ``trough'' at 4650--4690~\AA. See, for example, \citet{Wesemael1993}, their \S10 and Figure~16. The hottest PG~1159 stars can also exhibit the \ovi\ doublet in emission. In a classification scheme introduced by \citet{Werner1992}, the PG~1159 stars are subdivided into three groups according to the appearance of the \heii+\civ\ blend: ``A'' (showing only absorption), ``E'' (a mix of absorption with emission cores), and ``lgE'' (low-gravity emission; having relatively narrow absorption profiles underlying emission lines).

}

\section{Five PG~1159 Central Stars}

At this writing, we have obtained LRS2-B survey spectra for about four dozen faint PNNi. The present paper focuses on the six of them that have strong \ovi\ emission. To our knowledge, they have not been recognized previously as \ovi\ PNNi (although one was already known to belong to the PG~1159 class). Table~\ref{tab:targetlist} lists these objects and their PN~G designations, along with celestial and Galactic coordinates, and apparent $G$ magnitudes and $BP-RP$ colours, all taken from the recent \Gaia\/ Data Release~3\footnote{\url{https://vizier.cds.unistra.fr/viz-bin/VizieR-3?-source=I/355/gaiadr3}} (DR3; \citealt{Gaia2016, Gaia2022}). Our spectral types are given in the final column of Table~\ref{tab:targetlist}, and are discussed below. Details of our LRS2-B exposures are presented in Table~\ref{tab:exposures}.

\begin{table*}
\centering
\caption{Six ``\ovi'' Planetary Nuclei.}
\label{tab:targetlist}
\begin{tabular}{lcccccccc} 
	\hline
Name & PN G & RA (J2000) & Dec (J2000) & $l$  &  $b$  &     $G$  &   $BP-RP$ & Sp.\ Type \\
	\hline
Ou 2     & 120.4$-$01.3 &00 30 56.753 &  +61 24 34.29 &  120.48 & $-01.36$  &19.27 & $+0.36 $ & PG\,1159 (lgE) \\
Kn 61    & 070.5+11.0 & 19 21 38.934 &  +38 18 57.22 &  70.52  & $+11.01$  &18.24 & $-0.36 $ &  PG\,1159 (lgE) \\
Kn 15    & 064.5+03.4 & 19 40 40.337 &  +29 30 10.43 &  64.52  & $+03.41$  &17.69 & $+0.38 $ & [WO2] \\
Kn 12    & 060.3$-$05.0 & 20 03 22.544 &  +21 35 52.39 &  60.34  & $-05.03$  &18.44 & $-0.03 $ &  PG\,1159 (lgE) \\
Abell 72 & 059.7$-$18.7 & 20 50 02.052 &  +13 33 29.52 &  59.79  & $-18.73$  &16.00 & $-0.48 $ & PG\,1159 (lgE) \\
Kn 130   & 105.4$-$14.0 & 23 13 05.290 &  +45 26 18.29 &  105.42 & $-14.06$  &16.54 & $-0.32 $ &  PG\,1159 (lgE) \\
	\hline
\end{tabular}
\end{table*}

\begin{table}
\centering
\caption{HET LRS2-B Observations.}
\label{tab:exposures}
\begin{tabular}{lcc} 
Name &          Date    &       Exposure \\
     &     [YYYY-MM-DD] & [s]            \\
        \hline
Ou 2         &  2022-07-31   &   $2\times1200$ \\	      
Kn 61        &  2020-10-26   &   750	       \\
             &  2021-10-26   &   $2\times750$  \\ 	  
             &  2022-07-30   &   $2\times1000$ \\		  
Kn 15        &  2019-10-04   &   $2\times300$  \\
             &  2020-11-10   &   $2\times375$  \\  
             &  2021-11-10   &   $2\times375$  \\  
             &  2022-07-11   &   $2\times375$  \\ 				  
Kn 12        &  2019-10-06   &   $3\times500$  \\ 	  
             &  2021-10-30   &   $2\times1000$ \\  
             &  2022-07-10   &   $2\times1000$ \\				  
Abell 72     &  2019-08-05   &   180	       \\
             &  2020-08-06   &   150	       \\
             &  2021-08-01   &   450	       \\
             &  2021-11-07   &   150	       \\
             &  2022-05-15   &   360	       \\
Kn 130       &  2020-08-06   &   240	       \\
             &  2021-10-19   &   360	       \\
             &  2022-06-07   &   $2\times375$  \\ 	  
             &  2022-09-16   &   $2\times375$  \\
        \hline
\end{tabular}
\end{table}

\begin{figure*}
\centering
\includegraphics[width=5.75in]{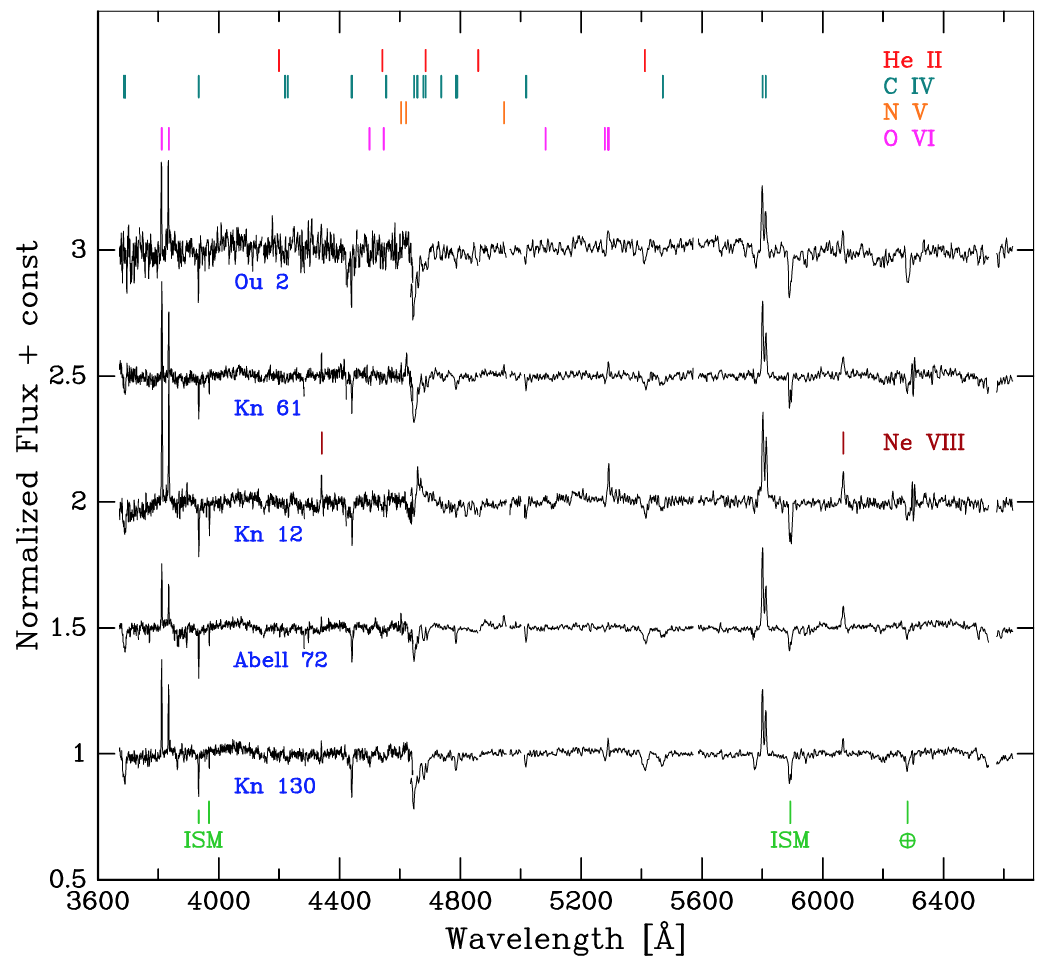}
\caption{HET/LRS2 spectra of five PG~1159 central stars showing \ovi\ emission. Tick marks on the vertical axis are spaced at half of the continuum level, and spectra have been offset successively by this amount. Wavelengths of prominent features are marked. Note strong emission at the \ovi\ 3811--3834~\AA\ and \civ\ 5801--5811~\AA\ doublets. Weak \nv\ emission is present at 4604, 4620, and 4945~\AA\ in several stars. \neviii\ emission lines are detected at 4340 in several objects, and are strong at 6068~\AA\ in all of them, implying very high effective temperatures (see text). Absorption features of \heii\ and \civ\ are conspicuous, typical of PG~1159 stars. Interstellar and telluric features are labelled at the bottom in green. Several instrument artifacts and inadequately subtracted nebular and night-sky emission lines have been suppressed. 
\label{fig:fivespectra} }
\end{figure*}

In one of the targets, Kn~15, the \ovi\ doublet is extraordinarily strong and blended; this remarkable object is described in the next section.
In the spectra of the other five objects, the \ovi\ emission lines are narrow and the doublet is well resolved.  In Figure~\ref{fig:fivespectra} we plot their LRS2 spectra. No significant time variability in the spectra was noted, and we plot equal-weight combinations of all of the exposures for each object. We have rectified the spectra to a flat continuum, and applied a 3-point boxcar smoothing (5-point for the faint Ou~2, for which we have only one HET visit), using standard tasks in {\tt IRAF}.\footnote{{\tt IRAF} was distributed by the National Optical Astronomy Observatories, operated by AURA, Inc., under cooperative agreement with the National Science Foundation.} Wavelengths of their strongest absorption and emission features are labelled at the top of the figure.\footnote{These line identifications come primarily from an unpublished list maintained by K.W., which draws significantly from the online Atomic Line List at \url{https://www.pa.uky.edu/~peter/atomic}; see \citet{vanHoof2018}.}

All five spectra in Figure~\ref{fig:fivespectra} are remarkably similar.\footnote{The apparent broad emission feature at $\sim$4660~\AA\ in Kn~12 is probably spurious; see \S\ref{subsec:kn12}.} In addition to the \ovi\ doublet, all five stars have strong emission at \civ\ 5801--5811~\AA\null. Kn~61 and Abell~72 exhibit \nv\ emission lines at 4604--4620 and 4945~\AA\null. Several of the spectra have an emission line at 4340~\AA, and all five have one at 6068~\AA; these features were identified with \neviii\ in a sample of PG~1159 stars by \citet{Werner2007}.\footnote{The wavelength of the \neviii\ 4340~\AA\ emission line is close to that of H$\gamma$. We verified that H$\gamma$ is very weak or absent in the sky spectra for the PNNi showing this line; thus the line at 4340~\AA\ did not result from under-subtraction of superposed nebular emission.} All five spectra have interstellar absorption at \caii\ 3933--3968~\AA\ and \nai\ 5889--5895~\AA, and telluric absorption at $\sim$6280~\AA, as labelled in green in the figure.

Numerous absorption lines are seen in the spectra. These are due to \heii, \civ, and \ovi, and are typical of the features seen in stars of the PG~1159 class. 
{
The 4650--4690~\AA\ absorption feature in all five stars shows emission cores; thus, in the final column of Table~\ref{tab:targetlist}, we assign all five stars a spectral classification of PG~1159~(lgE) (see discussion in \S4).
}

In the next five subsections we give details of each target and its surrounding PN\null. Our highest-SNR spectrum is that of the brightest target, Abell~72. In \S\ref{subsec:parameters} we compare it with a model-atmosphere synthetic spectrum, in order to make approximate estimates of its parameters and composition. Given the similarities of the five spectra, these parameters should roughly apply to all of the objects.


\subsection{Ou 2}

This faint PN was discovered by amateur N.~Outters during deep CCD imaging of a neighboring Galactic \hii\ region, as reported by \citet{Acker2012}. The nebula is roughly spherical, with a diameter of about $80''$. DSS images show a faint ($G=19.27$) blue star lying at the center, and our HET spectra are of this object. The star was also noted as a candidate WD or PNN during searches of \Gaia\/ data by \citet{GentileFusillo2019} and \citet{Chornay2020}, and was identified as the central star more recently by \citet{Ritter2022}. Nebular spectra obtained by \citet{Acker2012} and \citet{Ritter2022} show strong \heii\ 4686~\AA\ emission, implying a high temperature for the central star. 


\subsection{Kn 61}

Kn~61 was discovered by the DSH team \citep{Kronberger2012} as part of a targeted search for PNe in preparation for the {\it Kepler\/} mission. The PN has a diameter of about $100''$ and is approximately spherical, but with a filamentary structure; the discoverers proposed the name ``Soccer Ball Nebula.'' A deep image from the Gemini North telescope is given by \citet{DeMarco2015}. The $G=18.24$ central star was studied spectroscopically by \citet{Garcia-Diaz2014} and \citet{DeMarco2015}, and assigned a PG~1159 spectral type. Neither study obtained spectra at wavelengths short enough to cover the \ovi\ doublet; to our knowledge, we are the first to detect it in emission. 

\citet{Garcia-Diaz2014} reported photometric variability of the central star, with a peak-to-peak amplitude of $\sim$0.02~mag and a period of $5.7\pm0.4$~days. Because Kn~61 lies in the field in Cygnus that was monitored by the {\it Kepler\/} mission, \citet{DeMarco2015} were able to analyze the extensive {\it Kepler\/} photometry. They found unique behavior: there are sharp brightness peaks of $\sim$0.080 to 0.014~mag, occurring irregularly at intervals of 2 to 12~days, with durations of $\sim$1--2~days. They discussed the possible relation to the spectroscopic outbursts of Lo~4 \citep{Bond2014}, whose quiescent spectrum is very similar to that of Kn~61. However, the timescale of the events in Kn~61 is much shorter than in Lo~4, and it is not known whether Lo~4 exhibits photometric outbursts. In our spectroscopy of Kn~61, obtained at three widely separated epochs, we saw no significant changes in the spectrum. \citet{DeMarco2015} were unable to provide a definitive explanation for the remarkable photometric behavior. They did note that a WD with similar eruptive behavior exists: WD~J1916+3938 \citep{Hermes2011, Bell2015}. Moreover, \citet{Bell2016} found two more outbursting WDs in their examination of WDs in the {\it Kepler\/} field, and yet another case was found during the extended {\it Kepler\/} mission by \citet{Hermes2015}. But all of these erupting WDs are hydrogen-rich DA stars, having effective temperatures of $\sim$11,000~K, much cooler than the central star of Kn~61; thus the phenomena may be unrelated.


\subsection{Kn 12 \label{subsec:kn12} }

Kn~12 was classified as a possible PN in an early publication by the DSH team \citep{Kronberger2006}. A deep H$\alpha$ image is given by \citet{Jacoby2010}, showing a clumpy nebula with dimensions of about $54''\times 47''$; these authors called Kn~12 a probable PN\null. Our inspection of DSS images showed a conspicuous blue star near the center, which was also noted as a candidate WD or hot subluminous star in searches of \Gaia\/ data by both \citet{GentileFusillo2019} and \citet{Geier2019}.


Our spectrum of the Kn~12 central star shown in Figure~\ref{fig:fivespectra} appears to show a broad emission feature at $\sim$4660~\AA, in the vicinity of several lines of \civ\null. This feature is likely spurious, as it lies at the junction of the UV and Orange channels; at this location, the LRS2 system throughput varies steeply with wavelength, and moreover the transmission function of the dichroic beamsplitter is temperature-sensitive. Only two out of our three visits to Kn~12 showed this apparent emission feature.


\subsection{Abell 72}

This PN was discovered in the classical search of the Palomar Observatory Sky Survey (POSS) for faint nebulae by \citet{Abell1966}, who noted its 16th-magnitude central star. The PN has a diameter of $\sim\!155''$, with a filamentary structure whose similarity to the Soccer Ball Nebula Kn~61 is noted in the HASH catalogue. There are numerous studies of the nebula in the literature, but to our knowledge the only published spectroscopic study of the central star is by \citet{DeMarco2013}, who simply state that an ``Ultraviolet Visual Echelle Spectrograph spectrum suggests a temperature in excess of 120,000~K.'' 




\subsection{Kn 130}

This faint PN (size $277''\times 173''$) was discovered by the DSH team and announced by M.\,Kronberger et al.\ in an unpublished poster presentation at the Asymmetric Planetary Nebulae VII conference in 2017. As these authors noted, there is a conspicuous blue star near its center, which was our spectroscopic target. The star was also noted by \citet{Geier2019} in their search for hot subluminous stars in the \Gaia\/ catalogue. 




\subsection{Atmospheric Parameters and Compositions \label{subsec:parameters} }

In this subsection, we make an exploratory exploration of the atmospheric parameters and chemical compositions for these five stars, which, as noted above, all have very similar spectra.

The presence of emission lines from the 3s--3p doublets of \civ, \nv, and \ovi\ at 5801--5012, 4604--4620, and 3811--3834~\AA, respectively, in our five PG~1159 central stars implies effective temperatures well in excess of 100,000~K\null. As a typical example, in Figure~\ref{fig:abell72_model} we compare the spectrum of Abell~72 with that of a non-LTE model-atmosphere synthetic spectrum having $\Teff = 150,000$~K\null. Note that the model includes no stellar wind; the emission doublets of \civ, \nv, and \ovi\ arise in the photosphere. The model is taken from the grid introduced by \citet{werner14}, in which the T\"ubingen Model-Atmosphere Package {\citep[TMAP;][]{wernertmad2003} was used to compute non-LTE plane-parallel models in radiative and hydrostatic equilibrium. The compositions of the model atmospheres include He, C, N, and O\null. They do not include Ne, although it is detected in all five stars. The wings of the \heii\ absorption lines in our five targets indicate that their surface gravities are of  order $\log g \simeq 6.5$. 

\begin{figure*}
\centering
\includegraphics[width=5.75in]{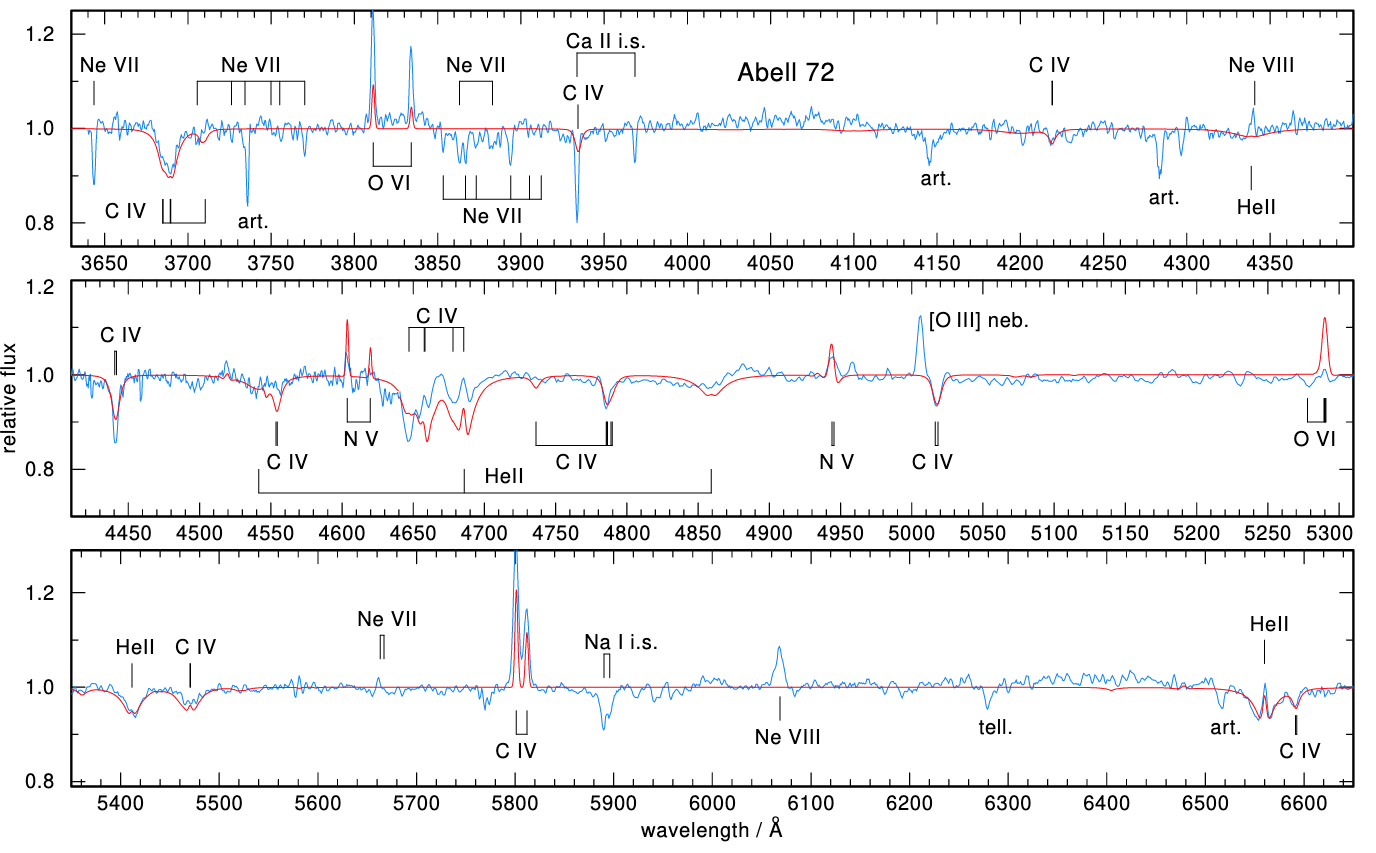}
\caption{Observed spectrum of the central star of Abell~72 (blue line) compared with a non-LTE model synthetic spectrum (red line) with $\Teff = 150,000$\,K and $\log g = 6.5$. The model includes helium, carbon, nitrogen, and oxygen, with mass fractions of He = 0.494, C = 0.490, N = 0.01, and O = 0.005. Labels indicate photospheric lines, interstellar absorption lines (i.s.), nebular emission lines (neb.), telluric absorption (tell.), and instrumental artifacts (art.). Note that there is no stellar wind in the model; the emission lines are photospheric.
\label{fig:abell72_model} }
\end{figure*}

As noted in the Figure~\ref{fig:abell72_model} caption, the atmosphere of Abell~72 is found to be composed almost entirely of helium and carbon. The oxygen abundance is difficult to estimate from the current spectra; however, it could be as low as $\sim$1\% by mass, because otherwise we would see broad and shallow absorption wings accompanying the \ovi\ emission line at 5290~\AA\ \citep{WernerHeberHunger1991}. The oxygen abundance in PG~1159 stars is often found to be significantly higher (of the order 10\%) than it is in Abell~72 (and the other four stars); however, similarly low abundances have been found in a few objects in previous studies \citep[e.g.,][]{wernerrauch2014,Lobling2019}. As usual in a number of PG~1159 stars, the presence of \nv\ lines points to a nitrogen abundance of the order of 1\% by mass \citep{WernerHerwig2006}. Although not included in the model, the same likely holds for neon, for which we identify a number of \nevii\ lines in Abell~72 and the other central stars. The \neviii\ emission lines at 4340 and 6068~\AA\ indicate that this star, and the other four, have effective temperatures even higher than that of the model, perhaps close to 170,000~K \citep{Werner2007}.


\section{The [WO2] Nucleus of K\lowercase{n}~15}

\subsection{Spectrum}

Kn~15 was noted as a possible PN in a visual survey of POSS plates by \citet{Roman2000}. The DSH team \citep{Kronberger2006} likewise catalogued it as a candidate PN, as did \citet{Viironen2009} in a search of the INT Photometric H-alpha Survey (IPHAS) catalogue. \citet{Jacoby2010} considered it a probable PN, presenting an H$\alpha$ image, and giving dimensions of $30''\times 23''$. The central star is a conspicuous 17th-magnitude blue star.

\begin{figure*}
\centering
\includegraphics[width=5.75in]{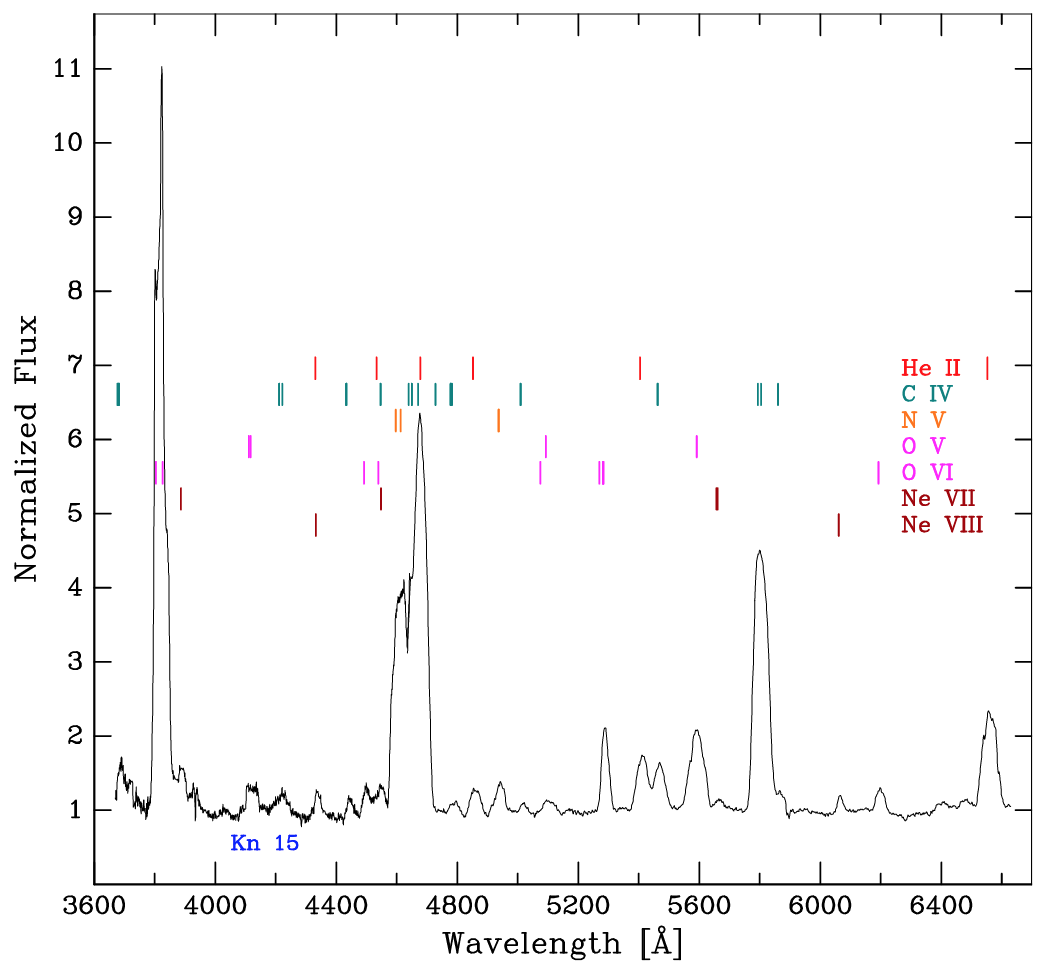}
\caption{Mean HET/LRS2 spectrum of the [WO2] central star of Kn~15 from four observation epochs, normalized to a flat continuum, with 3-point smoothing. Wavelengths of prominent emission features are marked. Note extraordinarily strong emission at the \ovi\ 3811--3834~\AA\ doublet, and conspicuous emission lines of \heii, \civ, \ov, and, remarkably, \nv\ (see text). Several lines of \nevii\ and \neviii\ are also present.
\label{fig:kn15spectrum} }
\end{figure*}

Figure~\ref{fig:kn15spectrum} plots our mean spectrum of the nucleus of Kn~15. This is an equally weighted combination of our eight exposures obtained over four epochs, normalized to a flat continuum, with 3-point boxcar smoothing. The spectrum shows an extremely strong and blended \ovi\ 3811--3834~\AA\ doublet, as well as prominent emission features of \heii\ and \civ\null. \ov\ 5590~\AA\ emission is strong, as is \nv\ 4604--4620~\AA\null. Weaker emission due to \ion{Ne}{VII-VIII} is also present.

This spectrum is similar to that of the star Sanduleak~3 (Sand~3), whose very strong \ovi\ emission was discovered five decades ago on an objective-prism plate by \citet{Sanduleak1971}. Until recently, Sand~3 was not known to be surrounded by an optically detectable PN\null. However, \citet{Gvaramadze2020} pointed out that mid-infrared images from the {\it Wide-field Infrared Survey Explorer\/} ({\it WISE}) show a bright clumpy nebula with a diameter of $\sim$$1'$, surrounded by a much fainter circular halo with a diameter of $6'$. These authors also detected several hydrogen-deficient optical knots lying within the clumpy mid-IR nebula. The 22~$\mu$m {\it WISE\/} image of Kn~15 (available at the HASH website) shows that it is likewise surrounded by a conspicuous mid-IR nebula, with hints of a faint circular outer halo similar to that of Sand~3. It would be of interest to determine whether there are hydrogen-deficient knots within the Kn~15 PN, in addition to the filamentary H$\alpha$ structures seen in the image presented by \citet{Jacoby2010}. 

Optical and UV spectroscopy of Sand~3 was discussed by \citet{Barlow1980} and \citet{BarlowHummer1982}. Although its spectrum superficially resembles that of a massive WO-type Wolf-Rayet star, Sand~3 lies at a Galactic latitude of $+12^\circ$; \citet{BarlowHummer1982} considered Sand~3 to be a low-mass star related to the central stars of PNe. This is now borne out by the relatively faint absolute magnitude implied by the \Gaia\/ parallax. In the classification scheme of \citet[][their Figure~7]{Crowther1998}, Sand~3 is assigned a spectral type of [WO1], the brackets denoting a low-mass PNN rather than a massive star. Our spectrum of the nucleus of Kn~15 indicates that it is slightly later in type (e.g., \ov\ 5590~\AA\ is stronger relative to \ovi\ than in Sand~3), and we classify it [WO2].   

The spectrum of Kn~15 is also similar in many regards to that of the [WO1] central star of the PN NGC~6905, recently discussed in detail by \citet{Gomez2022} (see especially their Figure~2), except that, again, \ov\ is stronger in Kn~15. 
However, the presence of emission lines of nitrogen in Kn~15, in particular the very strong \nv\ 4604--4620~\AA\ doublet, is remarkable. These lines are normally absent in hot Wolf-Rayet central stars, including NGC~6905, or very weak, as in Sand~3 \citep{Koesterke1997} or the [WC8] central star of NGC~40 \citep{Toala2019}. This points to a diversity of nitrogen abundances in [WO] and [WC] central stars---similar to what is observed in PG~1159 stars \citep[see][]{WernerHerwig2006}. 





\subsection{Spectral Variability}


To investigate whether the spectrum of Kn~15 varies with time, we intercompared our individual observations, obtained at four widely separated epochs between 2019 and 2022 (see Table~\ref{tab:exposures}). We find evidence that the spectrum does vary. In Figure~\ref{fig:kn15variations} we plot the spectra from the four epochs in the region between 5200 and 6000~\AA, which contains several strong emission lines of \heii, \civ, and \ovvi, and a weak line of \nevii\null. We see significant variations in the equivalent widths of the \civ\ 5801--5811~\AA\ blend and \ov\ 5590~\AA, by factors of $\sim$1.5 and 1.4, respectively. However, the \ovi\ 5290~\AA\ line is essentially non-variable. This is also true of the strong \ovi\ 3811--3834~\AA\ doublet in the UV channel (not plotted).  
By examining the spectra in absolute flux units, we verified that it is the emission lines that are varying; the continuum levels change by much smaller amounts, if at all.


\begin{figure}
\centering
\includegraphics[width=0.47\textwidth]{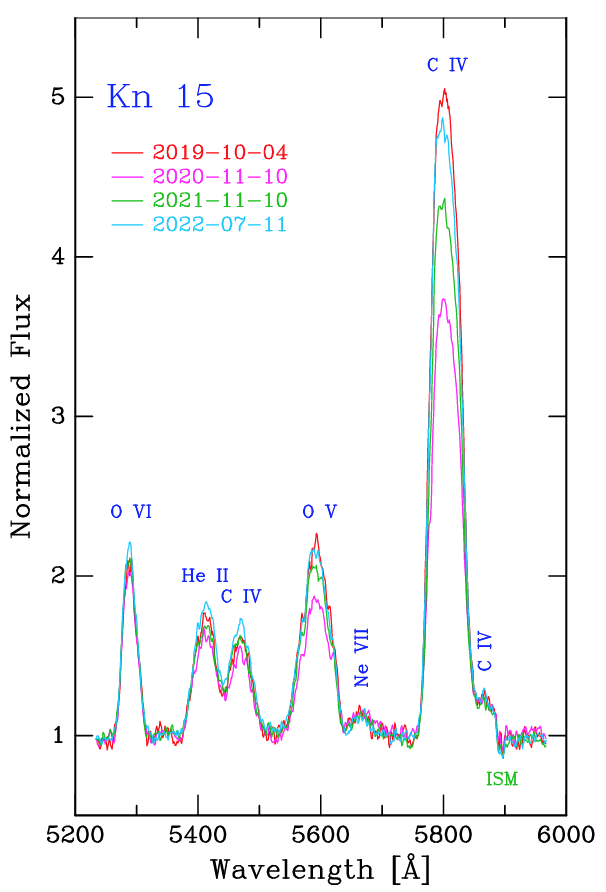}
\caption{Spectrum variations of Kn 15. We plot the 5200--6000~\AA\ region for our four epochs of HET/LRS2 observations. As in Figure~\ref{fig:kn15spectrum} the spectra are normalized to a flat continuum, and 3-point smoothing has been applied. Some of the emission lines, including \civ\ 5801--5811~\AA\ and \ov\ 5590~\AA, vary in equivalent width by factors of as much as $\sim$1.5, but others are nearly constant, as discussed in the text. 
\label{fig:kn15variations} }
\end{figure}

Variations in emission-line profiles in PNNi have been noted previously \citep[e.g.,][]{Balick1996,Grosdidier2000,Grosdidier2001}. However, these have generally been seen in late-type [WC] PNNi, such as the central star of NGC~40, and they have amplitudes of only a few percent.  \citet{Barlow1980} reported evidence for time variability of line profiles in their spectra of Sand~3, but the changes were likewise considerably more subtle than those we see in Kn~15. To our knowledge, emission-line equivalent-width changes by a factor of $\sim$1.5 have not been seen previously in a PNN, with the exception of the unique spectroscopic outbursts of the \ovi/PG~1159 nucleus of Lo~4 \citep{Bond2014}. They are indicative of variations or clumpiness in the stellar-wind outflow.

Our only constraint on the rapidity of the variations in Kn~15 comes from the two individual exposures that we obtained on four nights. These were typically spaced 6~minutes apart; none of them showed significant variations over that brief time span.


\section{Conclusions and Future Work}

We present initial results from an ongoing spectroscopic survey of central stars of faint PNe. The six stars discussed here all show strong emission at the \ovi\ doublet at 3811--3834~\AA, indicative of very high effective temperatures.

Five of the \ovi\ stars---the nuclei of Ou~2, Kn~61, Kn~15, Abell~72, and Kn~130---have similar spectra. They belong to the PG\,1159 spectral class, exhibiting strong absorption lines of \heii\ and \civ\null. In an exploratory examination of the data, we compare Abell~72 with a synthetic model-atmosphere spectrum. We conclude that its effective temperature (and by inference that of our other four PG\,1159 stars) is of order 170,000~K, as indicated by the presence of \neviii\ emission lines. Their atmospheres are composed of roughly equal amounts of helium and carbon, with small amounts ($\sim$0.5--1\%) of nitrogen, oxygen, and neon. 

The sixth star, the nucleus of Kn~15, has a spectrum resembling those of massive Wolf-Rayet stars, with strong and broad emission at the \ovi\ doublet, as well as conspicuous emission due to \heii, \civ, \nv, and \ov\null. The spectrum is similar to those of the [WO] central stars of Sand~3 and NGC~6905. We assign a spectral type of [WO2] to Kn~15, but we note that \nv\ is strikingly strong, indicating a relatively nigh nitrogen abundance. We urge spectroscopy of all six stars at higher spectral resolution, for detailed model-atmosphere analyses.

By comparing spectra of Kn~15 taken at four different epochs, we see variations in the equivalent widths of some emission lines by factors of as much as 1.5. These spectrum variations are the largest seen in any PNN, to our knowledge (other than the rare outbursts of the \ovi/PG\,1159 nucleus of Lo~4), and deserve detailed monitoring.

Sand 3, the central star of NGC~6905, and several other early-type [WO] PNNi are multiperiodic pulsating variables \citep[e.g.,][]{Bond1991,Ciardullo1996}; thus time-domain photometry of the Kn~15 nucleus would be of interest.

Similarly, about one-third of planetary nuclei and isolated WDs that have PG\,1159 spectral types have been found to be pulsating variables, belonging to the GW~Vir class \citep[e.g.,][]{Ciardullo1996, WernerHerwig2006, Althaus2010}. For example, both of the first two pulsating PNNi to be discovered, \hbox{K~1-16} \citep{Grauer1984} and Lo~4 \citep{BondMeakes1990}, have PG~1159 spectra with the \ovi\ doublet in emission \citep[e.g.,][]{Stanghellini1994,Bond2014}. Hence high-time-resolution photometry of our five PG\,1159 stars would be useful.


\section*{Acknowledgements}


We thank the HET queue schedulers and nighttime observers at McDonald Observatory for obtaining the data discussed here.

The Hobby-Eberly Telescope (HET) is a joint project of the University of Texas at Austin, the Pennsylvania State University, Ludwig-Maximilians-Universität M\"unchen, and Georg-August-Universit\"at G\"ottingen. The HET is named in honor of its principal benefactors, William P. Hobby and Robert E. Eberly.

The Low-Resolution Spectrograph 2 (LRS2) was developed and funded by the University of Texas at Austin McDonald Observatory and Department of Astronomy, and by Pennsylvania State University. We thank the Leibniz-Institut f\"ur Astrophysik Potsdam (AIP) and the Institut f\"ur Astrophysik G\"ottingen (IAG) for their contributions to the construction of the integral-field units.

We acknowledge the Texas Advanced Computing Center (TACC) at The University of Texas at Austin for providing high-performance computing, visualization, and storage resources that have contributed to the results reported within this paper.

This work has made use of data from the European Space Agency (ESA) mission
{\it Gaia\/} (\url{https://www.cosmos.esa.int/gaia}), processed by the {\it Gaia\/}
Data Processing and Analysis Consortium (DPAC,
\url{https://www.cosmos.esa.int/web/gaia/dpac/consortium}). Funding for the DPAC
has been provided by national institutions, in particular the institutions
participating in the {\it Gaia\/} Multilateral Agreement.

\section*{Data Availability}




The individual calibrated spectra are available at \url{https://doi.org/10.5281/zenodo.7613614}. 



\bibliographystyle{mnras}
\bibliography{ovi_pnni_refs} 








\bsp	
\label{lastpage}
\end{document}